\pdfoutput=1 
\documentclass[letterpaper,english,reprint, aps]{revtex4-1}
\usepackage[T1]{fontenc}
\usepackage[latin9]{inputenc}
\usepackage{verbatim}
\usepackage{amsmath}
\let\oldAA\AA
\renewcommand{\AA}{\text{\normalfont\oldAA}}\usepackage{amssymb}
\usepackage{graphicx}
\PassOptionsToPackage{version=3}{mhchem}
\usepackage{mhchem}

\makeatletter

\providecommand{\tabularnewline}{\\}

\usepackage{physics}

\makeatother

\usepackage{babel}
\begin{document}
\preprint{APS/123-QED}
\title{An efficient direct band-gap transition in germanium by three-dimensional
strain}
\author{S. Mellaerts\textsuperscript{1}, V. Afanasiev\textsuperscript{1},
J.W. Seo\textsuperscript{2}, M. Houssa\textsuperscript{1,3} and
J.-P. Locquet\textsuperscript{1}}
\affiliation{\textsuperscript{1}Department of Physics and Astronomy, KU Leuven,
Celestijnenlaan 200D, 3001 Leuven, Belgium, \textsuperscript{2}Department
of Materials Engineering, KU Leuven, Kasteelpark Arenberg 44, 3001
Leuven, Belgium, \textsuperscript{3}Imec, Kapeldreef 75, 3001 Leuven,
Belgium}
\email{simon.mellaerts@kuleuven.be}

\begin{abstract}

Complementary to the development of highly three-dimensional (3D)
integrated circuits in the continuation of Moore's law, there has
been a growing interest in new 3D deformation strategies to improve
device performance. To continue this search for new 3D deformation
techniques, it is essential to explore beforehand - using computational
predictive methods - which strain tensor leads to the desired properties.
In this work, we study germanium ($\ce{\ce{Ge }}$) under an isotropic
3D strain on the basis of first-principle methods. The transport and
optical properties are studied by a fully \emph{ab initio} Boltzmann
transport equation and many-body Bethe-Salpeter equation (BSE) approach,
respectively. Our findings show that a direct band gap in $\ce{\ce{Ge }}$
could be realized with only $0.34\%$ triaxial tensile strain (negative
pressure) and without the challenges associated with $\ce{\ce{Sn }}$
doping. At the same time a significant increase in refractive index
and carrier mobility - particularly for electrons - is observed. These
results demonstrate that there is a huge potential in exploring the
3D deformation space for semiconductors - and potentially many other
materials - in order to optimize their properties.
\end{abstract}
\maketitle

\section{\label{sec:Intro}Introduction}

The advances in nanotechnology over the past decade have led to the
continuation of Moore's law in the third dimension. As two-dimensional
(2D) scaling of transistors is reaching its limit on various levels,
there has been a development of 3D integration techniques which involves
the vertical stacking and connecting of multiple functional materials
and devices \citep{3Dintegrationcircuits,3Dintegration_IBM}. This
has led to an industry paradigm shift to explore the various benefits
such an approach offers, including small form factor, multifunctionality
as well as increased performance and yield.

This development also opens up a rich avenue for the exploration of
epitaxially strained 3D materials. Epitaxial strain engineering is
widely adopted in tuning and enhancing the functional properties of
various materials including semiconductors \citep{StrainedSi_roadmap,StrainEffect_Sun,StrainstudyGe}
and oxide thin films \citep{JP_doublingTc,JP_strainV2O3,Piapaper}.
However, in most cases, epitaxy involves the introduction of an in-plane
strain by a lattice mismatch with the substrate, whereas the out-of-plane
component is free to relax in order to to sustain the induced stress
in a nearly volume conserving manner. Furthermore, the lateral strain
in thin films is limited by a critical thickness, above which the
strain will slowly relax through the gradual introduction of misfit
dislocations. A very recent approach promises to fundamentally alter
this traditional picture. Through the engineering of vertically aligned
nanocomposite (VAN) thin films it becomes possible to maintain the
strain to much higher thickness by coupling to the vertical interfaces
\citep{3DstrainVAN,VAN_nature,VAN_aip}. Once an epitaxial coupling
to vertical interfaces can be realized, 3D tensile strain becomes
envisionable.

In similar deviations from conventional epitaxy, there have been several
attempts to access crystal phases deviating from their equilibrium
crystal symmetry by the use of transfer methods. One example is the
prediction and experimental realization of hexagonal $\ce{\ce{Si }}$,
$\ce{\ce{Ge }}$ and $\ce{\ce{Si_{1-x}Ge_{x} }}$ nanowires through
crystal transfer methods in which a wurtzite gallium phosphide ($\ce{\ce{GaP }}$)
nanowire core is used as template \citep{DFT-hexSiGe,hex-Si_exprealization,exprealization_hexSiGe,hexSiGe_nature}.
Furthermore, it was shown that hex-$\ce{\ce{Si_{1-x}Ge_{x} }}$ nanowires
can have a direct band gap offering great opportunities for group
III-V semiconductor opto-electronic devices.

\ 

In light of these new 3D material deformation strategies, we explore
here the properties of 3D isotropically strained $\ce{\ce{Ge }}$.
It has been shown recently that $\ce{\ce{GeSn }}$ alloys grown on
$\ce{\ce{Si }}$ can also exhibit a direct band gap - depending on
the amount of Sn doping - which is investigated in the context of
both laser \citep{GeSn_NL-EPM,GeSn_VCA-MBJ,GeSn_IncreasedPL,GeSn_lasingOnSi,GeSnLasing_nature}
and electronic applications \citep{GeSn_DFTbands,GeSn_mobility,GeSn_ruben}.
On the other hand, a direct band gap has already been observed in
$\langle111\rangle$ uniaxial and $(001)$ biaxial tensile strained
$\ce{\ce{Ge }}$ at values $\sim4\%$ ($E_{g}=0.40$ eV) and $\sim2\%$
($E_{g}=0.39$ eV), respectively \citep{UniaxialGe,GeBiaxial,StrainstudyGe}.
Note that in such deformations, the cubic symmetry is lost which is
detrimental for the valley degeneracies. In this work we demonstrate
using an \emph{ab initio} study that a symmetry conserving tensile
strain above $0.34\%$ in all three directions would be sufficient
to induce a much larger direct band gap ($716$ meV) and circumventing
the need for $\ce{\ce{Sn }}$ doping in Ge.

\section{\label{sec:Methods}Computational methods}

All calculations were performed within density functional theory (DFT)
as implemented in the Vienna \emph{ab initio} simulation package (VASP)
\citep{VASP}. The interactions between electrons and ions were described
by the projector augmented wave (PAW) potentials \citep{PAWmethod},
and the electronic wave functions were expanded with a cutoff energy
of $500$ eV. To study the various strained configurations, the choice
of the exchange-correlation functional for the lattice and electronic
band structure was optimized with respect to respectively the experimental
lattice constant and energy gap of the unstrained configuration. For
the ionic relaxation, phonon and elastic calculations the local density
approximation (LDA) implemented by Ceperley-Alder was adopted \citep{LDA}.
For the structural relaxation, a force convergence criterion of $0.005$
eV/$\AA$ was used with the Brillouin zone (BZ) sampled by a
$12\times12\times12$ $\Gamma$-centered $k$-scheme. Phonon dispersions
were calculated self-consistently on the basis of density functional
perturbation theory (DFPT) and with the use of the PHONOPY package
\citep{phonopy}, while the stiffness matrix was calculated by a finite
difference method as implemented in VASP.

For the electronic calculations, the improved hybrid functional for
solids HSEsol \citep{HSEsol} was chosen as it gave the best experimental
agreement for the energy gap. The BZ was sampled by a $16\times16\times16$
$\Gamma$-centered $k$-scheme with energy convergence criterion of
$10^{-6}$ eV.

To accurately describe the optical properties a fully self-consistent
many-body approach was adopted \citep{GWBSE}. As a starting point,
the HSEsol calculated wave functions were used as input for the self-consistent
Green's function (GW) method. Subsequently, these Green's functions
are used to solve the Bethe-Salpeter equations (BSE) \citep{BSE1,BSE2},
which take into account the excitonic effects. Due to the large computational
expense of the BSE spectrum a less dense $8\times8\times8$ $\Gamma$-centered
$k$-scheme is adopted.

An estimate of the transport properties was obtained by the use of
the AMSET package. This approach completely relies on first-principe
input parameters where the scattering rates are calculated by the
Boltzmann transport equation (BTE) within the momentum relaxation
time approximation (MRTA). To account for the band shifts and band
warping the (anisotropic) deformation potentials were calculated via
AMSET, resulting in relatively accurate mobility calculations at low
computational expense \citep{AMSET}.

\section{\label{sec:Results}Results and Discussion}

\subsection{\label{subsec:Structural-and-elastic}Structural and elastic properties}

\begin{figure}
\includegraphics[scale=0.95]{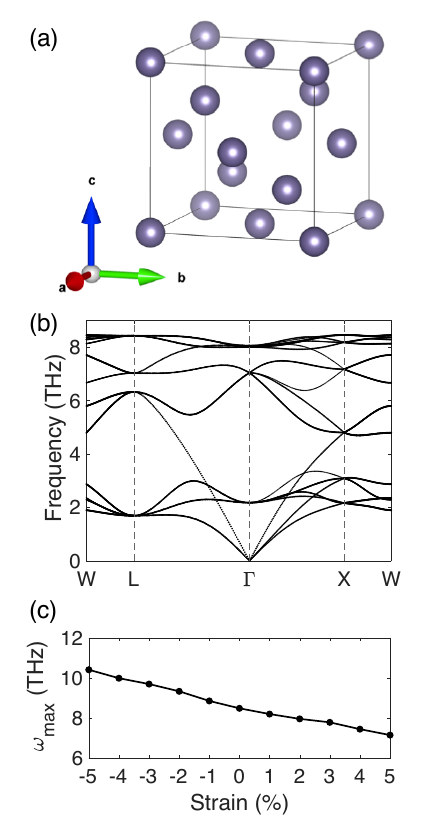}\caption{Lattice structure of germanium and its dynamical properties. (a) Conventional
fcc unit cell. (b) Calculated phonon dispersion and (c) maximum phonon
frequency as a function of 3D strain.}
\label{fig:Structural}

\end{figure}

The structurally optimized face-centered cubic (fcc) unit cell with
space group $Fd\bar{3}m$ (no. $227)$ is shown in Fig. \ref{fig:Structural}.
The lattice constant equals $5.65\, \AA$, which is in agreement
with the experimental value. An isotropic 3D strain $\varepsilon$,
defined as $\varepsilon=(a-a_{0})/a_{0}$, was applied along all three
cubic lattice vectors within a range of $-5\%$ to $+5\%$ and stepsize
$1\%$. Positive strain corresponds to a tensile expansion. For each
strained configuration all atomic degrees were optimized. It was found
that the space group is preserved and the $\ce{\ce{Ge }}$-$\ce{\ce{Ge }}$
bond length changes linearly by the application of the isotropic strain.

By symmetry, the $6\times6$ stiffness matrix only consists of three
independent stiffness constants $C_{11},C_{12}$ and $C_{66}$. From
the finite difference method, the stiffness constants were calculated,
from which the respective elastic parameters were derived: the bulk
modulus ($B$), shear modulus ($G$), and Young's modulus ($Y_{s}$).
These derived elastic parameters, which are summarized in Table \ref{tab:Elastic},
are within $7\%$ of the experimental parameters.

\begin{figure*}
\includegraphics[scale=0.65]{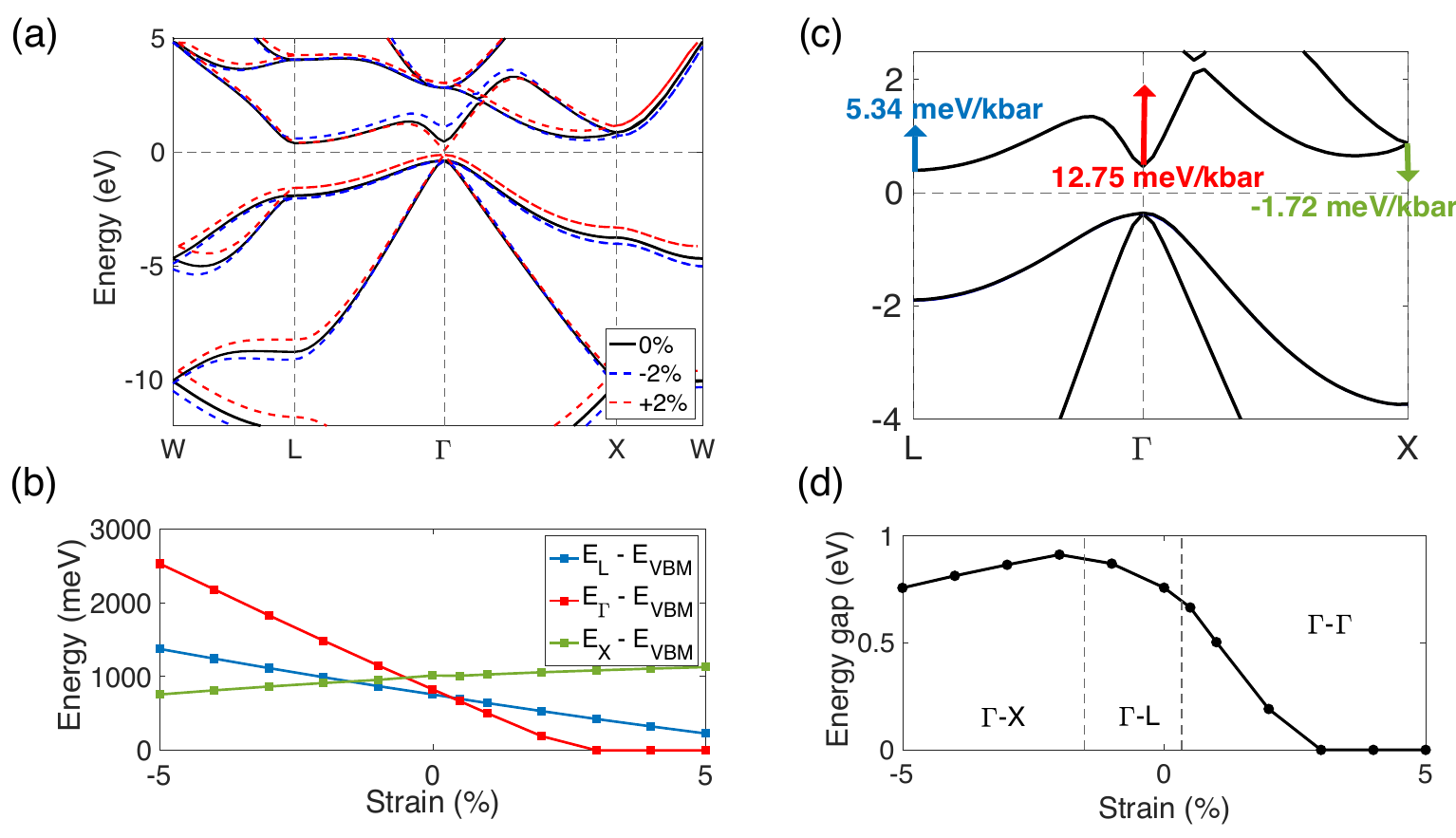}\caption{Electronic structure as a function of 3D strain. (a) The band structure
as a function of strain and (b) the strain dependences of the band
edges with respect to the VBM at $\Gamma$ and (c) the derived pressure
dependence of these band edges at the three BZ points {[}$\mathrm{L}$,
$\Gamma$, $\mathrm{X}${]}. (d) The evolution of the band gap as
it evolves between 3 different CBM points in the BZ .}
\label{fig:Electronic}
\end{figure*}

The dynamical properties of the equilibrium and strained configurations
were determined by calculating the phonon dispersion. Our main quantity
of interest here, is the cutoff phonon frequency $\omega_{max}$ which
gives the upper bound for electron-phonon coupling interactions. It
is shown in Fig. \ref{fig:Structural}c that this cutoff frequency
shows a linear dependence on hydrostatic pressure, where it decreases
under 3D tensile strain. A similar trend has been observed in $\ce{\ce{Si }}$
\citep{DielectricBreakdown_si}.

\begin{table}[h]
\caption{The DFT calculated stiffness constants and the derived elastic parameters
compared to the experimental (Exp) values taken from \citep{ExpElastic}.}
\label{tab:Elastic}

\begin{tabular}{|c|c|c|c|c|c|c|}
\hline 
 & $C_{11}$ & $C_{12}$ & $C_{66}$ & $B$ & $G$ & $Y_{s}$\tabularnewline
\hline 
\hline 
DFT (kbar) & $1197$ & $475$ & $600$ & $716$ & $361$ & $927$\tabularnewline
\hline 
Exp (kbar) & $1260$ & $480$ & $670$ & $740$ & $390$ & $995$\tabularnewline
\hline 
\end{tabular}
\end{table}

\subsection{\label{subsec:Electronic-band-structure}Electronic band structure}

To accurately describe the electronic band structure and to obtain
experimental agreement with the energy gap, the improved hybrid functional
for solids HSEsol was employed. The resulting band structure as a
function of strain is shown in Fig. \ref{fig:Electronic}a. In the
equilibrium situation the band structure of $\ce{\ce{Ge }}$ has an
indirect gap of $757$ meV with its valence band maximum (VBM) located
at the $\Gamma$ point and conduction band minimum (CBM) at $\mathrm{L}$.
By evaluating the evolution of the band edges relative to VBM at $\Gamma$
throughout the strain range (see Fig. \ref{fig:Electronic}b), the
pressure dependence of the band edges was determined at three points
in the BZ, as depicted in Fig. \ref{fig:Electronic}c. This agrees
well with earlier predictions \citep{GeSn_bandmixing}.

From the evolution of the band edges and the energy gap, shown in
Fig. \ref{fig:Electronic}b-d, it is clear that the strong pressure
dependence of the band edges allows two band gap transitions; from
a compressive strain above $1.52\%$ the CBM changes from the $\mathrm{L}$
to $\textrm{X}$ point, while a tensile strain of $0.34\%$ induces
a direct gap transition at $\Gamma$ with an estimated gap value of
$716$ meV. On further increasing the tensile strain to $2.5\%$ the
direct gap at the $\Gamma$ point will be closed and $\ce{\ce{Ge }}$
becomes metallic, as derived from linear fitting of the $\Gamma$
edge evolution in Fig. \ref{fig:Electronic}b. In contrast to the
$\sim4\%$ tensile $\langle111\rangle$ uniaxial ($E_{g}=0.40$ eV)
or $\sim2\%$ tensile $(001)$ biaxial strain ($E_{g}=0.39$ eV) inducing
a direct band gap \citep{StrainstudyGe}, the 3D strain does not affect
the degeneracy of the valleys, as symmetry is preserved. Moreover,
from Fig. \ref{fig:Electronic}d, it is clear that the maximum direct
band gap value that can be achieved is much larger compared to earlier
approaches \citep{GeSn_DFTbands,DFT_uni-biaxialstrain,DFT-hexSiGe},
while the 3D strain also allows the band gap to be tuned within this
enlarged range of $0-716$ meV.

\begin{figure*}
\includegraphics[scale=0.62]{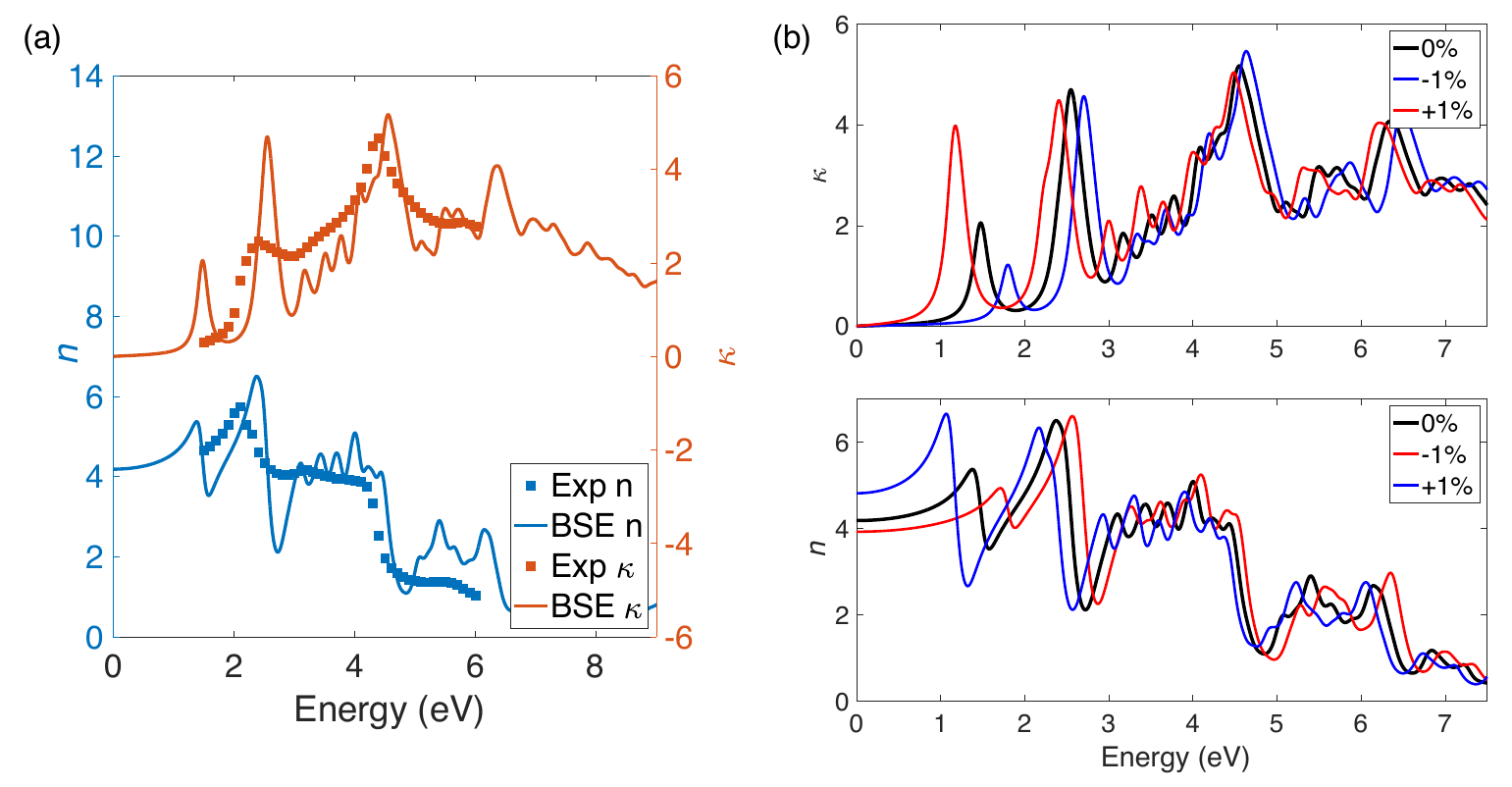}\caption{Optical properties as a function of 3D strain. (a) The GW-BSE calculated
energy-dependent refractive index $n$ and extinction coefficient
$\kappa$ compared to the experimental (Exp) data taken from \citep{ExpOptics}.
(b) The strain dependence of $\kappa$ and $n$ .}
\label{fig:Optical}
\end{figure*}

\begin{table}[h]
\caption{The derived effective masses as a function of 3D strain.}
\label{Table:EffMass}

\begin{tabular}{|c|c|c|c|c|c|c|c|c|}
\hline 
Strain (\%) & $-5$ & $-4$ & $-3$ & $-2$ & $-1$ & $0$ & $+1$ & $+2$\tabularnewline
\hline 
\hline 
$m_{e}^{*}/m_{0}$ & $0.846$ & $0.821$ & $0.835$ & $0.851$ & $0.129$ & $0.110$ & $0.035$ & $0.032$\tabularnewline
\hline 
\hline 
$m_{hh}^{*}/m_{0}$ & $0.202$ & $0.208$ & $0.216$ & $0.223$ & $0.229$ & $0.241$ & $0.242$ & $0.251$\tabularnewline
\hline 
$m_{lh}^{*}/m_{0}$ & $0.091$ & $0.086$ & $0.083$ & $0.077$ & $0.072$ & $0.066$ & $0.061$ & $0.060$\tabularnewline
\hline 
\end{tabular}
\end{table}

Furthermore, the isotropic electron ($m_{e}^{*}$) and light/heavy
hole ($m_{lh}^{*}/m_{hh}^{*}$) effective masses were derived from
the curvature of the CBM and VBM valley, respectively, summarized
in Table \ref{Table:EffMass}. The induced hydrostatic pressure has
relative little effect on the effective masses of the light and heavy
holes in the $\Gamma$ valley, nevertheless, the variation in position
of the CBM valley is accompanied by drastic changes in the effective
mass of the conduction electrons. Therefore, the realization of the
direct gap at $0.34\%$ tensile strain comes with the additional benefit
of an extremely high decrease in the electron effective mass $m_{e}^{*}$
at the CBM. 

\subsection{\label{subsec:Optical-properties}Optical properties}

The establishment of a direct band gap whose value can be carefully
tuned by the 3D strain opens the possibility of many optical applications.
Therefore, the frequency-dependent dielectric function has been calculated
by solving the BSE, taking into account excitonic effects. From the
calculated complex dielectric function $\varepsilon=\varepsilon_{1}+i\varepsilon_{2}$,
the refractive index $n$ and extinction coefficient $\kappa$ were
determined by

\begin{align}
n= & \frac{1}{\sqrt{2}}\left(\varepsilon_{1}+(\varepsilon_{1}^{2}+\varepsilon_{2}^{2})^{1/2}\right)^{1/2},\\
\kappa= & \frac{1}{\sqrt{2}}\left(-\varepsilon_{1}+(\varepsilon_{1}^{2}+\varepsilon_{2}^{2})^{1/2}\right)^{1/2}.
\end{align}

The resulting optical parameters are shown in Fig. \ref{fig:Optical}a
and reproduce well the main features of the experimental data \citep{ExpOptics}.
Therefore, this self-consistent GW-BSE approach provides a good estimate
of the optical properties, although it should be emphasized that the
self-consistency in the GW calculation and the HSEsol wave functions
as input were crucial to obtain reliable results. Subsequently, the
dielectric function and derived optical parameters for the different
strained configurations were determined, whose trend in the tensile
and compressive regime are depicted in Fig. \ref{fig:Optical}b. Furthermore,
an estimate of the static dielectric constant was obtained by the
low-frequency asymptote of $\varepsilon_{1}$. The static dielectric
constants ($\varepsilon_{0}$) and corresponding refractive index
($n$) are summarized in Table \ref{Table:Dielectric}.

\begin{table}[h]
\caption{The derived static dielectric constant ($\varepsilon_{0}$) and refractive
index ($n$) as a function of 3D strain.}
\label{Table:Dielectric}

\begin{tabular}{|c|c|c|c|c|c|c|c|c|}
\hline 
Strain (\%) & $-5$ & $-4$ & $-3$ & $-2$ & $-1$ & $0$ & $+1$ & $+2$\tabularnewline
\hline 
\hline 
$\varepsilon_{0}$ & $12.57$ & $12.96$ & $13.53$ & $14.26$ & $15.41$ & $17.51$ & $23.21$ & $39.36$\tabularnewline
\hline 
$n$ & $3.55$ & $3.60$ & $3.68$ & $3.78$ & $3.93$ & $4.18$ & $4.82$ & $6.27$\tabularnewline
\hline 
\end{tabular}
\end{table}

\begin{figure*}
\includegraphics[scale=0.67]{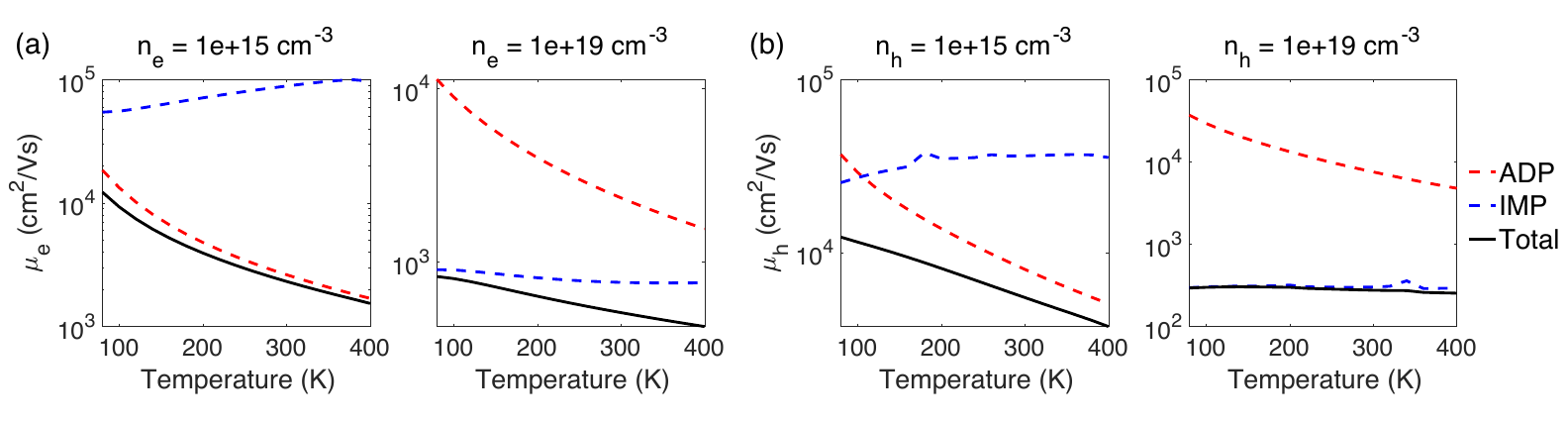}\caption{The (a) electron and (b) hole mobility as a function of doping concentration
($n_{e}$/$n_{h}$), and the respective contributions from acoustic
deformation potential (ADP) scattering and impurity (IMP) scattering.}
\label{fig:MobilityScattering}
\end{figure*}

The transition to the direct band gap at $1\%$ tensile strain causes
a significant increase in the interband transition probability across
the energy gap. This manifests itself in a drastic increase of the
static dielectric permittivity and the redshift of the first peak
in the extinction coefficient - proportional to the absorption coefficient
- with a corresponding increase of factor $2$. Since the symmetry
and VBM degeneracy are preserved under 3D strain, the system becomes
very similar to $\ce{\ce{GaAs }}$ but with a much smaller energy
gap. This offers great opportunities for infrared photodetectors \citep{ptypeGaAs_photodetector,GaAs_QWdetector}.
Alternatively, a small anisotropy in the 3D strain could be applied
to induce LH and HH band splitting, where the LH band forms a low
density of states (DOS) VBM, which facilitates population inversion,
lowering the lasing threshold \citep{GeSn_lasingNature}. In respect
of achieving population inversion, the fraction of conduction electrons
populating the $\Gamma$ valley ($N_{\Gamma}/N_{CB}$) at an injection
carrier density $N_{inj}$ of $10^{19}\:\textrm{cm}^{-3}$, which
is a reasonable charge carrier density in the lasing regime \citep{SuggestionCalculation_Valeri},
is calculated. By evaluation of the DOS with assuming a thermal broadening
of $300$ K, $N_{\Gamma}/N_{CB}$ is found to be $4\%$ and $9\%$
for tensile strain of $0.5\%$ ($E_{\Gamma}-E_{\mathrm{L}}=32$ meV)
and $1\%$ ($E_{\Gamma}-E_{\mathrm{L}}=137$ meV), respectively. These
$N_{\Gamma}/N_{CB}$ values are similar to what has been reported
for the achieved $E_{\Gamma}-E_{\mathrm{L}}$ in $\ce{\ce{GeSn }}$
alloys, while a much larger $N_{\Gamma}/N_{CB}$ fraction of $27\%$
- twice the largest reported fraction \citep{SuggestionCalculation_Valeri}
- can be achieved by a $2\%$ triaxial tensile strain due to an extremely
large valley difference $E_{\Gamma}-E_{\mathrm{L}}=340$ meV.

\subsection{\label{subsec:Transport-properties}Transport properties }

The fully \emph{ab initio }Boltzmann transport equation was solved
within the MRTA while the scattering rates were evaluated by Fermi's
Golden rule. As the fcc crystal structure of $\ce{\ce{Ge }}$ is centrosymmetric
the main scattering mechanisms consist of acoustic phonon scattering
and impurity scattering, which are assumed to be elastic processes.
The former scattering mechanism is implemented by an advanced acoustic
deformation potential (ADP) scattering formalism that includes the
scattering from both longitudinal and transverse modes in a single
matrix element \citep{AMSET}. Therefore, a fully anisotropic deformation
potential matrix was constructed by calculating the electronic structure
under three inequivalent deformations. On the other hand, the quantum
mechanical treatment of Brooks and Herring \citep{Brooks,Herring}
was adopted for approximating the impurity (IMP) scattering rate where
the screening length was evaluated by the calculated DOS and static
dielectric permittivity. Subsequently, the temperature-dependent electron/hole
mobility ($\mu$) was derived from the scattering rate and effective
mass.

\ 

In this way, both temperature-dependent electron and hole mobility
were calculated for the unstrained case as a function of the doping
concentration, depicted in Fig. \ref{fig:MobilityScattering}. At
low doping concentrations, the scattering rate is dominated by interactions
with acoustic phonons, where carriers experience both intra- and intervalley
scattering. At room temperature, the latter process becoming the main
scattering mechanism due to the degenerate nature of the $\mathrm{L}$
valley. Therefore, uni- and biaxial strain have been successful in
the enhancement of the electron mobility by lifting this degeneracy
\citep{DFT_uni-biaxialstrain,tensileGe_NW}. While at higher doping
concentrations, it is the impurity scattering that limits both electron
and hole mobility.

\ 

\begin{figure}
\includegraphics[scale=0.67]{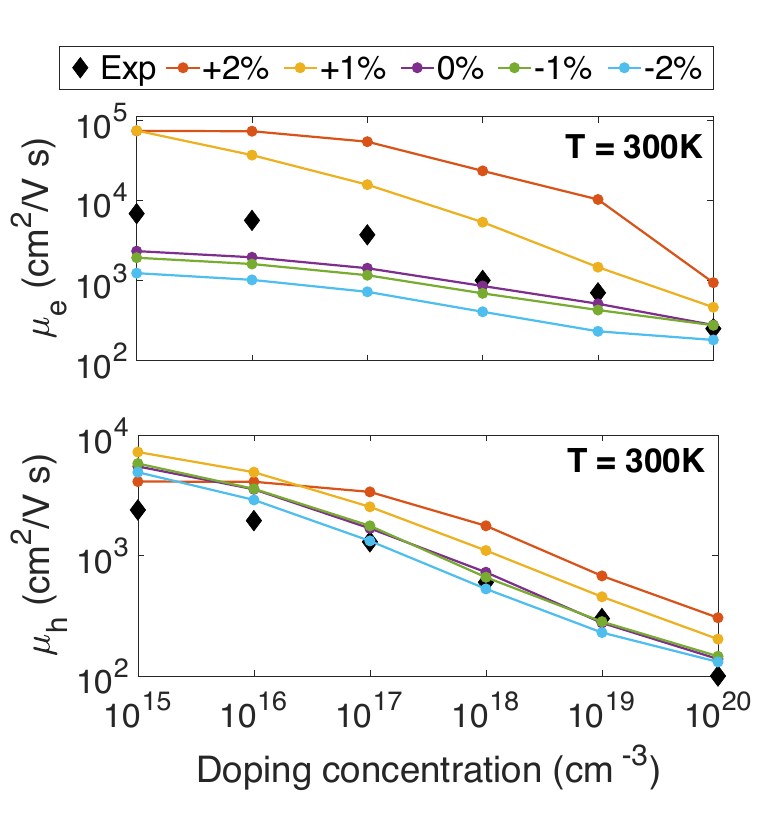}\caption{The room-temperature electron (top) and hole (bottom) mobility as
a function of doping concentration for various strained configurations,
compared to experimental data taken from \citep{Ge_expEdoping,Ge_expHdoping}.}
\label{fig:Mobility}
\end{figure}

Subsequently, the room-temperature electron and hole mobility were
evaluated as a function of 3D strain and compared to the experimental
data, shown in Fig. \ref{fig:Mobility}. At high doping concentrations,
the simulated mobilities agree very well with the experimental data,
whereas at lower doping concentrations the agreement is still within
standard computational deviations. These deviations can be expected
as the IMP scattering is much easier to calculate than ADP, since
the latter strongly depends on small variations in the band energies.

In the compressive region, the electron mobility reduces slightly
when the nature of the indirect gap changes with the CBM located at
the $\textrm{X}$ point. Conversely, the electron mobility increases
significantly under $1-2\%$ tensile strain, which can be mostly attributed
to CBM located at the $\Gamma$ with a reduced effective mass. This
increased electron mobility is significant in the low doping regime
where ADP scattering dominates, with a maximum enhancement of the
electron mobility by a factor of $20-50$ compared to pure $\ce{\ce{Ge }}$
($3900\:\textrm{cm}^{2}/\textrm{V\ensuremath{\cdot}s}$) at doping
concentration $10^{15}\,\textrm{cm}^{-3}$. The increase is less pronounced
in the high doping regime where IMP scattering limits the mobility.
In this latter situation, the enhancement is the result of the improved
screening by an increased static dielectric permittivity $\varepsilon_{0}$. 

On the other hand, the induced hydrostatic pressure by 3D strain has
a smaller effect on the hole mobility as the VBM degeneracy is preserved
and the hole effective masses change only slightly. The hole mobility
is nearly constant in the compressive region, while an increase of
$50-100\%$ is observed for $1-2\%$ tensile train in the high doping
regime which can again be explained by the increased static dielectric
permittivity $\varepsilon_{0}$. 

\section{Discussion}

To fully appreciate the potential of these results, a proper comparison
to $\ce{\ce{Sn }}$ alloying and uni- and biaxial strain should be
made. Earlier computational work already pointed out that the perpendicular
relaxation to the tensile strained direction is unfavorable to induce
a direct gap transition. A direct gap transition without perpendicular
relaxation could be achieved at $2.7\%$ $[111]$ uni- and $1\%$
$(001)$ biaxial strain, while due to perpendicular relaxation an
additional $1\%$ and $0.5\%$ strain respectively is required \citep{StrainstudyGe}.
From this perspective, it is not surprising to observe a direct gap
transition at $0.34\%$ tensile strain in all three spatial directions.
In following, the advantages in the low and high band gap regime in
electronic and optical applications are discussed.

Our first-principle calculations suggest that this 3D strain offers
great opportunities for $\ce{\ce{Ge }}$ high-mobility transistors
with several advantages over current approaches. The enhancement in
electron mobility by a factor of $20-50$ compared to pure $\ce{\ce{Ge }}$
at doping concentration $10^{15}\,\textrm{cm}^{-3}$ is quantitatively
very similar to maximum mobilities achieved in highly uni- and biaxial
strained $\ce{\ce{Ge }}$ and $\ce{\ce{GeSn }}$ alloys \citep{DFT_uni-biaxialstrain,GeSn_mobility}.
This can be expected as they all rely on the inducement of a direct
gap where the mobility is determined by the conduction electrons populating
the $\Gamma$ valley with a reduced effective mass. Nonetheless, we
expect that the enhancement of the electron mobility should also be
more substantial for 3D strain induced lower band gap states compared
GeSn systems where an increasing $\ce{\ce{Sn }}$ alloy disorder scattering
becomes detrimental for the electrons populating the $\Gamma$ valley
\citep{GeSn_mobility}.

In addition, our results suggest that these current approaches are
inefficient in achieving a direct band gap as their maximum possible
direct gap is relatively low. In contrast to 3D strain with a maximum
direct gap value of $0.72$ eV, a much lower value of $0.40$ eV and
$0.39$ eV is found for uni- and biaxial strain respectively \citep{StrainstudyGe},
and a similar value of $0.48$ eV for $\ce{\ce{GeSn }}$ alloys \citep{GeSn_DFTbands}.
These lower band gap values impose strong constraints on the performance
of high-mobility devices.

\ 

This also has implications for optical devices, where a much larger
direct gap value range ($0-716$ meV) can be covered by the 3D strain
within an experimental accessible strain range of $\sim2.5\%$, achievable
by the use of $\ce{\ce{In_{x}Ga_{1-x}As }}$ buffer layers \citep{tensileGe_optics}.
This becomes particularly interesting in the search for group III-V
semiconductor opto-electronic applications where lower band gap values
are required to cover the complete infrared region (e.g. gas sensing
at $5\:\mu\textrm{m}$/$248$ meV). Achieving these lower gap values
in $\ce{\ce{GeSn }}$ alloys would require very high $\ce{\ce{Sn }}$
concentrations, which quickly becomes experimentally challenging due
to frequent $\ce{\ce{Sn }}$ precipitation \citep{GeSn_growth}. Furthermore,
random alloy fluctuations and band mixing lead to inhomogeneous broadening
of the band edges and gain spectrum, resulting in the increase of
the lasing threshold \citep{GeSn_bandmixing}. An other major concern
is the relatively low DOS of the CBM which puts constraints on the
density of electrons that can be excited to the $\Gamma$ valley before
saturation and spill over to the $\mathrm{L}$ valley. Achieving a
direct gap nature at much higher gap value by 3D strain implies an
increased $E_{\Gamma}-E_{\mathrm{L}}$ and corresponding increase
of the fraction of conduction electrons populating the $\Gamma$ valley
($N_{\Gamma}/N_{CB}$) at the required band gap value for infrared
lasing. The current $\ce{\ce{GeSn }}$ lasers are operating with band
gap values of $460$ meV and $E_{\Gamma}-E_{\mathrm{L}}<70$ meV \citep{GeSnLasing_nature,GeSn_lasingNature},
whereas by 3D strain of $1.1\%$ the same direct gap could be achieved
with an increased valley energy difference of $164$ meV. These differences
will increase further in the achievement of higher wavelength lasers,
which again shows the advantages of efficiently inducing a direct
band gap in $\ce{\ce{Ge }}$ through 3D strain (negative pressure).

Although this remains experimentally unexplored for $\ce{\ce{Ge }}$,
we suggest that the inducement of 3D strain might - besides using
VAN's - also be achieved by the gradual introduction of strain through
a thermal expansion mismatch with the substrate. In this way, the
epitaxial deposition of $\ce{\ce{Ge }}$ in a $\ce{\ce{Si }}$ micrometer
trench would allow a tensile strain of $0.25\%$ in all three directions
\citep{Liu_GeonSi,Liu_lasingGeonSi}. By interpolation of our data,
this would result in a nearly direct band gap with only an energy
difference of $E_{\mathrm{L}}-E_{\Gamma}=17$ meV. This is in contrast
to $0.25\%$ biaxially strained $\ce{\ce{Ge }}$ on $\ce{\ce{Si }}$
with $E_{\mathrm{L}}-E_{\Gamma}=115$ meV \citep{Liu_lasingGeonSi},
which has been used as a lasing platform, However, in this situation
a high n-doping concentration of $\sim10^{19}\:\textrm{cm}^{-3}$
is required to compensate for the indirect gap nature and to fill
the $\mathrm{L}$ valley. Therefore, our proposed scheme could circumvent
the need of this extremely high n-doping which is accompanied by large
free carrier absorption loss and an increasing lasing threshold.

\section{\label{sec:Conclusion}Conclusions}

Our first-principle simulations predict that 3D strain offers a very
interesting and efficient alternative route for the realization of
a direct band gap in single crystalline $\ce{\ce{Ge }}$. This has
several advantages including a precise control of the band gap value
which allows coverage of the whole infrared emission region within
the experimental feasible strain range of $\sim2.5\%$ , essential
for optical implementations. In this context, 3D strain prevents various
disadvantages of the $\ce{\ce{Sn }}$ alloying of $\ce{\ce{Ge }}$
method where random alloy fluctuations limit the electron mobility
at lower band gap values, and band mixing effects lead to inhomogeneous
broadening of the band edges and gain spectrum, increasing the lasing
threshold \citep{GeSn_bandmixing}. We hope these findings will stimulate
the search for new 3D deformations techniques as an alternative to
current approaches. As a future step, also anisotropic 3D strains
should be considered since these will for instance enable to control
the existence of the degeneracies in the respective valleys.

\section*{Acknowledgements}
Part of this work was financially supported by the KU Leuven Research Funds, Project No. KAC24/18/056 and No. C14/17/080 as well as the Research Funds of the INTERREG-E-TEST Project (EMR113) and INTERREG-VL-NL-ETPATHFINDER Project (0559). Part of the computational resources and services used in this work were provided by the VSC (Flemish Supercomputer Center) funded by the Research Foundation Flanders (FWO) and the Flemish government.
\bibliographystyle{unsrt}
\bibliography{final-Manuscript.bib}

\end{document}